\documentclass{ieeeaccess}
\usepackage{cite}
\usepackage{amsmath,amssymb,amsfonts}
\usepackage{algorithmic}
\usepackage{graphicx}
\usepackage{textcomp}
\usepackage{mathtools}
\usepackage{multirow}
\usepackage{url}
\usepackage{color}
\usepackage{setspace}
\usepackage{booktabs}
\usepackage{subfigure}
\usepackage{braket}
\usepackage{framed}
\usepackage{bm}

\def\BibTeX{{\rm B\kern-.05em{\sc i\kern-.025em b}\kern-.08em
    T\kern-.1667em\lower.7ex\hbox{E}\kern-.125emX}}

\begin{document}
\history{Date of publication xxxx 00, 0000, date of current version xxxx 00, 0000.}
\doi{10.1109/ACCESS.2017.DOI}

\onecolumn
\begin{framed}
   \noindent
   This work has been submitted to the IEEE for possible publication. Copyright may be transferred without notice, after which this version may no longer be accessible.%
\end{framed}
\clearpage

\title{Evaluating the Performance of Direct Higher-Order Formulations in Combinatorial Optimization Problems}
\author{\uppercase{Kazuki Ikeuchi}\authorrefmark{1},
\uppercase{Yoshiki Matsuda\authorrefmark{2}, and Shu Tanaka}\authorrefmark{1,3,4,5},
\IEEEmembership{Member, IEEE}}
\address[1]{Graduate School of Science and Technology, Keio University, 3-14-1 Hiyoshi, Kohoku-ku, Yokohama-shi, Kanagawa 223--8522, Japan}
\address[2]{Fixstars Corporation, 3-1-1, Shibaura, Minato-ku, Tokyo 108--0023, Japan}
\address[3]{Department of Applied Physics and Physico-Informatics, Keio University, Kanagawa 223--8522, Japan}
\address[4]{Keio University Sustainable Quantum Artificial Intelligence Center (KSQAIC), Keio University, Tokyo 108--8345, Japan}
\address[5]{Human Biology-Microbiome-Quantum Research Center (WPI-Bio2Q), Keio University, Tokyo 108--8345, Japan}

\tfootnote
    {This work was partially supported by the Council for Science, Technology, and Innovation (CSTI) through the Cross-ministerial Strategic Innovation Promotion Program (SIP), ``Promoting the application of advanced quantum technology platforms to social issues'' (Funding agency: QST), the Japan Society for the Promotion of Science (JSPS) KAKENHI (Grant Number JP23H05447), Japan Science and Technology Agency (JST) (Grant Number JPMJPF2221).}

\markboth
{Kazuki Ikeuchi \headeretal: 
Evaluating the Performance of Direct Higher-Order Formulations in Combinatorial Optimization Problems}
{Kazuki Ikeuchi \headeretal: 
Evaluating the Performance of Direct Higher-Order Formulations in Combinatorial Optimization Problems}

\corresp{Corresponding author: Kazuki Ikeuchi (e-mail: kazuki.ike.1028@keio.jp).}

\begin{abstract}
    Ising machines, including quantum annealing machines, are promising next-generation computers for combinatorial optimization problems. 
    However, due to hardware limitations, most Ising-type hardware can only solve objective functions expressed in linear or quadratic terms of binary variables. 
    Therefore, problems with higher-order terms require an order-reduction process, which increases the number of variables and constraints and may degrade solution quality.
    In this study, we evaluate the effectiveness of directly solving such problems without order reduction by using a high-performance simulated annealing-based optimization solver capable of handling polynomial unconstrained binary optimization (PUBO) formulations.
    We compare its performance against a conventional quadratic unconstrained binary optimization (QUBO) solver on the same hardware platform. 
    As benchmarks, we use the low autocorrelation binary sequence (LABS) problem and the vehicle routing problem with distance balancing, both of which naturally include higher-order interactions.
    Results show that the PUBO solver consistently achieves superior solution quality and stability compared to its QUBO counterpart, while maintaining comparable computational time and requiring no order-reduction compilation indicating potential advantages of directly handling higher-order terms in practical optimization problems.
\end{abstract}


\titlepgskip=-15pt

\maketitle

\section{Introduction}
\label{sec:introduction}

\PARstart{C}{ombinatorial} optimization problems are widely encountered in various fields of modern society, such as logistics~\cite{neukart2017traffic,bao2021approach,kanai2024annealing}, materials design~\cite{kitai2020designing,inoue2022towards,hashiguchi2025material}, and portfolio optimization~\cite{rosenberg2016solving,tanahashi2019application,hidaka2023correlation,tatsumura2023real}.
These problems are formulated as mathematical models comprising decision variables with combinatorial structures, objective functions, and constraints.
Let the decision variable vector with a combinatorial structure be denoted by $\bm{z}$, and its entire search space by $\mathcal{S}$.
Furthermore, the subset of feasible solutions that satisfy all constraints is denoted by $\mathcal{F}$.
The goal of combinatorial optimization is to find the decision variable vector that minimizes the objective function while satisfying all constraints, that is, 
\begin{equation}
    {\bm z}^* = \operatorname*{argmin}_{\bm z} f_{\operatorname*{obj}}({\bm z}),
    \quad {\bm z} \in {\mathcal F} \subseteq {\mathcal S},
\end{equation}
where $\bm{z}^*$ represents the optimal solution.

However, it is well known that the number of candidate solutions (i.e., the size of ${\mathcal S}$) increases exponentially with the number of decision variables, making exhaustive search computationally infeasible for large-scale problems.
Therefore, various heuristic and metaheuristic algorithms have been developed to obtain near-optimal solutions within a practical computational time.

Among the representative metaheuristic approaches, Simulated Annealing (SA) is a well-established method for combinatorial optimization.
SA interprets the objective function as an energy function and introduces a temperature parameter to control the stochastic search process, gradually lowering the temperature to explore low-energy states.
Thus, SA can be regarded as an algorithm based on state transitions driven by thermal fluctuations.

Inspired by the concept of SA, Quantum Annealing (QA) has been proposed as a quantum-mechanical extension~\cite{kadowaki1998quantum,tanaka2017quantum,chakrabarti2023quantum}.
In QA, quantum fluctuations play a role analogous to thermal fluctuations in SA.
Similar to SA, the objective function is treated as an energy function, and a parameter controlling the strength of quantum fluctuations is gradually decreased to obtain low-energy configurations. 

In recent years, hardware implementations of such physics-inspired algorithms, referred to as Ising machines, have been developed~\cite{oku2020reduce,mohseni2022ising,kikuchi2025effectiveness}.
An Ising machine is designed to search for low-energy states of an Ising model or its equivalent Quadratic Unconstrained Binary Optimization (QUBO) formulation.
In recent years, various algorithmic improvements have been developed to enhance the performance of Ising machines, such as hybrid quantum-classical schemes~\cite{feld2019hybrid, kitai2020designing, ding2021implementation, kikuchi2023hybrid, tamura2025black}, variable-reduction techniques~\cite{irie2021hybrid, shirai2023spin, ide2025extending}, and appropriate tuning of hyperparameters~\cite{tamura2021performance,shirai2020guiding,ayodele2022multi,kikuchi2023dynamical,hattori2025impact}.
In parallel, Ising machines have been applied to a wide range of practical problems in various fields~\cite{yarkoni2022quantum,yulianti2022implementation}.
Since most existing hardware platforms accept QUBO as the standard input format, the energy function must be expressed as a polynomial composed solely of linear and quadratic terms of binary variables.
Consequently, problems involving higher-order interactions require order-reduction techniques to transform Polynomial Unconstrained Binary Optimization (PUBO) problems into the QUBO form~\cite{hauke2020perspectives,schmidbauer2024polynomial}.

Several studies have been conducted to develop efficient methods for performing this order reduction~\cite{biamonte2008nonperturbative,babbush2013resource}, and such approaches have proven to be effective in solving various optimization problems involving higher-order terms.
However, order reduction techniques inevitably introduce additional auxiliary variables and constraints to maintain consistency between the original and transformed formulations.
These additional components increase the problem size and structural complexity, potentially degrading the overall solution quality.
Nevertheless, numerous real-world problems inherently involve higher-order interactions, such as polymer sampling~\cite{cristian2021polymer, francesco2023quantum}, protein folding~\cite{carlos2021investigating, chermoshentsev2022polynomial}, and optimizing molecular adsorption configurations on alloy surfaces~\cite{tuan2025optimizing}.
Accordingly, developing solvers capable of directly handling higher-order terms is becoming increasingly important.

Motivated by this background, this study compares two distinct approaches for solving combinatorial optimization problems, one that directly handles higher-order terms without order reduction, and another that applies order reduction techniques to convert problems into the QUBO form.
The comparative evaluation focuses on the solution quality and performance characteristics of both formulations. 

For this purpose, the Fixstars Amplify Annealing Engine (Amplify AE)~\cite{FixstarsAmplify} was employed. 
Amplify AE is a high-performance optimization solver based on the simulated annealing algorithm, designed to efficiently solve combinatorial optimization problems through energy-based searches on Ising models.
In addition to conventional quadratic optimization problems, Amplify AE provides a polynomial optimization solver that can directly handle up to fourth-order terms without order reduction. 
Since both solvers operate on the same hardware platform and share the same underlying algorithm, Amplify AE provides a fair and consistent environment for evaluating performance differences attributable solely to order reduction.

Previous studies have also investigated the implications of solving higher-order problems directly, rather than reducing them to a quadratic form. 
For instance, comparisons of energy gaps in the 3-SAT problem have shown that maintaining the PUBO formulation can be advantageous for adiabatic performance compared to the reduced QUBO form~\cite{nagies2025boosting}.
Similarly, studies employing the Quantum Approximate Optimization Algorithm (QAOA) on continuous optimization benchmarks have suggested that retaining the PUBO form can achieve comparable or superior accuracy with fewer required variables~\cite{stein2023evidence}.
Furthermore, comparative studies using simulated bifurcation machines have demonstrated performance differences between PUBO and QUBO formulations under identical physical principles~\cite{kanao2023simulated}. 
In addition, recent developments in gate-based quantum algorithms have explored direct encoding and utilization of higher-order interactions, indicating that the importance of directly solving PUBO problems extends beyond quantum annealing and quantum-inspired computing paradigms.

Although these previous studies have highlighted the significance of retaining higher-order terms, most of them have focused on artificial benchmarks or relatively small-scale instances.
In contrast, real-world applications often involve large-scale problems with complex constraints, where the impact of order reduction is expected to be more pronounced.
Based on this perspective, the originality of this study lies in quantitatively comparing PUBO and QUBO formulations on the same algorithmic foundation for practical combinatorial optimization problems and  assessing their respective solution accuracy and characteristics. 
Through this analysis, the study aims to extend previous theoretical insights and small-scale demonstrations, thereby revealing the potential advantages of directly solving PUBO formulations in realistic large-scale optimization scenarios.

The remainder of this paper is structured as follows.
Section~\ref{sec:background} provides an overview of Ising machines, their underlying mechanisms, and the order-reduction method.
In Section~\ref{sec:ourproblems}, we introduce two benchmark problems considered in this study.
The numerical results are presented in Section~\ref{sec:results}, followed by discussions in Section~\ref{sec:discussion}.
Lastly, we draw conclusions in Section~\ref{sec:conclusion}.
\section{Background}
\label{sec:background}

\subsection{Ising formulation and QUBO basics}
\label{sec:ising_formulation}
In this section, we describe the procedure for solving combinatorial optimization problems based on the Ising model.
As discussed in Section~\ref{sec:introduction}, combinatorial optimization problems appear in many domains of modern society, and many of them are known to be NP-hard, meaning that the computational cost of obtaining exact solutions grows exponentially with problem size. 
As a result, exhaustive search becomes impractical for large-scale instances.
Consequently, various approximation algorithms and metaheuristics have been developed over the years.
More recently, physics-inspired approaches have attracted attention as a new paradigm for tackling such problems.

A representative example of these approaches is the Ising machine, a hardware system designed to search for the ground state of the Ising model.
In certain cases, Ising machines have demonstrated performance surpassing that of conventional classical algorithms.
The energy function of the Ising model is expressed as
\begin{align}
\label{Eq:Isingmodel}
    E_{\mathrm{Ising}}({\bm s}) &= \sum_{i=1}^N h_i s_i + \sum_{1 \le i < j \le N} J_{i,j} s_i s_j, \\
    s_i &\in \{+1, -1\},
\end{align}
where $\bm s$ represents a set of $N$ spin variables $s_i$.
The first term in \eqref{Eq:Isingmodel} corresponds to the local field $h_i$ acting on spin $s_i$, and the second term represents the pairwise interaction $J_{i,j}$ between spins $s_i$ and $s_j$.
Here, both $h_i$ and $J_{i,j}$ are real constants.
Finding the ground state of this Hamiltonian corresponds to obtaining the optimal solution to the corresponding combinatorial optimization problem.

When a problem is expressed using binary variables $x_i \in \{0,1\}$, it can be formulated as a quadratic unconstrained binary optimization (QUBO) problem:
\begin{align}
    \label{Eq:QUBO}
    E_{\mathrm{QUBO}}({\bm x}) &= \sum_{1 \le i \le j \le N} Q_{i,j} x_i x_j, \\
    x_i &\in \{0,1\},
\end{align}
where $\bm x$ denotes the set of binary variables, and $Q_{i,j}$ are real-valued coefficients.
The Ising and QUBO formulations are mathematically equivalent through the linear transformation $s_i=2x_i-1$.
Accordingly, Ising-type hardware typically accepts QUBO-form energy functions as standard input.

\subsection{Simulated Annealing}
\label{sec:simulated_annealing}
Simulated Annealing (SA) is a stochastic optimization algorithm originally proposed by Kirkpatrick et al.~\cite{kirkpatrick1983optimization}.
Inspired by the annealing process in metallurgy, SA introduces a temperature parameter to probabilistically escape from local minima during the search for a global optimum.

In the context of Ising-type computation, the objective function is expressed in the Ising form
\begin{equation}
    E(\bm{s}) = -\sum_{i<j} J_{ij}s_is_j - \sum_i h_i s_i,
\end{equation}
or equivalently in the QUBO formulation.
SA iteratively updates the spin configuration $\bm{s}$ and accepts candidate states $\bm{s}'$ according to the Metropolis criterion, which depends on the energy difference $E(\bm{s}') - E(\bm{s})$:
\begin{equation}
P_{\mathrm{acc}}(\bm{s} \to \bm{s}') =
\begin{cases}
1, & \text{if } \Delta E \le 0, \\
\exp\!\left(-\dfrac{\Delta E}{k_\mathrm{B} T}\right), & \text{if } \Delta E > 0,
\end{cases}
\end{equation}
where $T$ denotes the temperature and $k_\mathrm{B}$ is the Boltzmann constant.
The temperature is gradually decreased according to a predefined cooling schedule, such as geometric cooling, expressed as
\begin{equation}
    T_{k+1} = \gamma T_k \quad (0 < \gamma < 1),
\end{equation}
which is commonly used in practice.

In this study, numerical experiments were conducted using Amplify AE, a high-performance simulated annealing solver based on large-scale GPU parallelization.
Amplify AE supports both PUBO and QUBO formulations, enabling direct handling of up to fourth-order terms.
Because both solvers share a common algorithmic foundation, performance differences between PUBO and QUBO can be fairly compared without the influence of order-reduction transformations involving auxiliary variables or penalty terms. 
This feature makes Amplify AE a suitable experimental platform for evaluating optimization problems that inherently include higher-order interactions.

As shown in \eqref{Eq:QUBO}, the QUBO formulation consists solely of linear and quadratic terms of binary variables. 
Therefore, problems that inherently involve third- or higher-order interactions cannot be directly represented in QUBO form.
However, as discussed in Section~\ref{sec:introduction}, many practical combinatorial optimization problems naturally contain higher-order terms.
To make such problems compatible with Ising-type hardware, order-reduction techniques are applied by introducing auxiliary variables and penalty terms that convert higher-order terms into equivalent quadratic forms.
The next subsection describes a representative order-reduction technique known as the substitution method.

\subsection{Order-reduction methods}
\label{sec:order_reduction}
In this section, we describe the order-reduction techniques used to convert higher-order polynomial terms into the QUBO form.
As discussed in Section~\ref{sec:ising_formulation}, the QUBO formulation includes linear and quadratic terms.
Therefore, when higher-order terms appear, they must be reduced to a quadratic form through order-reduction techniques.
Several methods have been proposed for reducing the order of higher-degree polynomial terms, such as the substitution method~\cite{biamonte2008nonperturbative} and the Ishikawa–Kolmogorov–Zabih–Freeman–Delong (Ishikawa–KZFD) method~\cite{ishikawa2010transformation,kolmogorov2004energy,delong2012fast,freedman2005energy}.
In this study, we employ the substitution method, which is widely used and mathematically straightforward.
A brief overview of the method is provided below.

The substitution method can mitigate the increase in the number of variables by efficiently reusing auxiliary variables.
In this approach, the product of two variables in higher-order terms is replaced with a newly introduced auxiliary variable, thereby reducing the order step by step.

For example, when a cubic term $x_1x_2x_3$ is given, the substitution method reduces it to quadratic form as follows:
\begin{align}
x_1x_2x_3 = y_{1,2}x_3 + p[x_1x_2 - 2y_{1,2}(x_1+x_2) + 3y_{1,2}].
\label{Eq:substitution}
\end{align}
In~\eqref{Eq:substitution}, the product $x_1x_2$ is replaced with a new auxiliary binary variable $y_{1,2}$, and a penalty term is introduced to enforce the substitution constraint.
\begin{table}
    \centering
    \caption{Relationship among $x_1,x_2,y_{1,2}$ and AND($x_1,x_2,y_{1,2}$)}
    \label{Tab:substitution}
    \begin{tabular}{cccc}
    \hline
    $x_1$ & $x_2$ & $y_{1,2}$ & AND($x_1,x_2,y_{1,2}$) \\
    \hline \hline
    0 & 0 & 0 & 0 \\
    1 & 0 & 0 & 0 \\
    0 & 1 & 0 & 0 \\
    1 & 1 & 0 & 1 \\
    0 & 0 & 1 & 3 \\
    1 & 0 & 1 & 1 \\
    0 & 1 & 1 & 1 \\
    1 & 1 & 1 & 0 \\
    \hline
    \end{tabular}
\end{table}
To illustrate this equation, Table~\ref{Tab:substitution} lists all possible combinations of $x_1$, $x_2$, and $y_{1,2}$, along with the corresponding value of the function $\mathrm{AND}(x_1, x_2, y_{1,2})$.
Here, we define
\begin{align}
\text{AND}(x_1,x_2,y_{1,2}) = x_1x_2-2y_{1,2}(x_1+x_2)+3y_{1,2}.
\end{align}
From Table~\ref{Tab:substitution}, it can be observed that AND($x_1,x_2,y_{1,2}$) takes positive values when $x_1x_2 \neq y_{1,2}$.
Therefore, the second term in \eqref{Eq:substitution} imposes a penalty when the substitution constraint $x_1x_2 = y_{1,2}$ is not satisfied.
Here, $p$ denotes the penalty coefficient that controls the strength of the substitution constraint, and the solution accuracy is sensitive to its value.

As mentioned at the beginning of this section, the substitution method can mitigate the increase in auxiliary variables compared to other order-reduction techniques by performing substitutions for multiple terms simultaneously.
Nevertheless, the total number of variables still grows rapidly with problem order, which can significantly affect the solution quality and computational stability.
In addition, the solution accuracy remains sensitive to the value of the penalty coefficient $p$, which controls the strength of the substitution constraint.
These aspects, namely the penalty setting and the variable expansion, are discussed in detail in Section~\ref{sec:discussion}.
\section{Our problems}
\label{sec:ourproblems}
In this section, we introduce two benchmark problems considered in this study. 
The first is the low autocorrelation binary sequence (LABS) problem, which serves as a representative benchmark in quantum optimization research and naturally involves fourth-order interactions in its formulation.
The second is the vehicle routing problem (VRP) with distance-balancing, described in Section~\ref{sec:vrp_prob}, which is a multi-objective optimization problem motivated by practical applications.
This problem aims to minimize the total travel distance while equalizing the travel distances among vehicles, leading to higher-order interactions among binary variables.
These two problems are employed to compare the performance of formulations that directly retain higher-order terms with those converted into QUBO form through order reduction.

\subsection{Low autocorrelation binary sequence problem}
\label{sec:labs_prob}
\begin{figure}
    \centering 
    \includegraphics[scale=0.4]{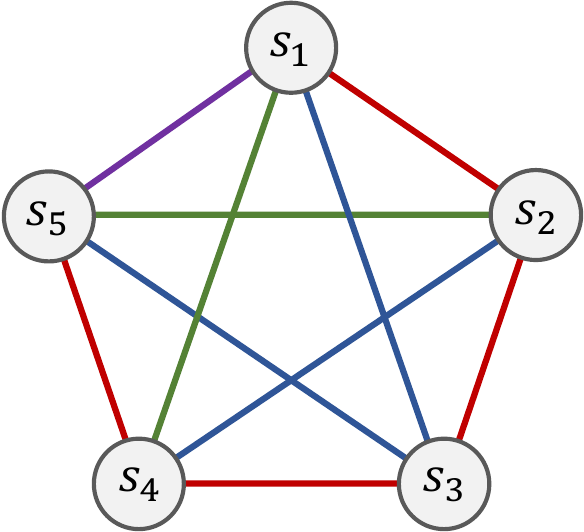}
    \caption{Conceptual illustration of the correlation structure in the LABS problem for $N=5$.
    Each node represents a spin variable $s_i \in \{-1, +1\}$.
    Edges of the same color correspond to spin pairs contributing to the autocorrelation term $C_k$ at lag $k$, where red, blue, green, and purple represent shifts of $k = 1, 2, 3,$ and $4$, respectively.
    The overall energy $E = \sum_{k=1}^{N-1} C_k^2$ is given by the sum of squared correlations at different lags, indicating that the interactions among spins arise from the squared combination of these color-coded pairwise correlations.}    
    \label{Fig:labsimage}
\end{figure}

The low autocorrelation binary sequence (LABS) problem~\cite{golay1977sieves,schroeder1970synthesis,hoholdt2002determination,bernasconi1987low} is an optimization problem that seeks a sequence of binary variables $\bm{s} = \{s_1, s_2, \dots, s_N \}$, each taking values $-1$ or $+1$, that minimizes the autocorrelation energy, as illustrated schematically in Fig.~\ref{Fig:labsimage}.
The autocorrelation corresponding to a shift $k$ is defined as
\begin{equation}
    C_k(\bm{s}) = \sum_{i=1}^{N-k} s_i s_{i+k}, \quad k=1, \dots, N-1, 
\end{equation}
and the total energy function is given by the sum of squared autocorrelations, 
\begin{equation}
    E(\bm{s}) = \sum_{k=1}^{N-1} C_k(\bm{s})^2.
\end{equation}

The LABS problem is important in various applications where sequences with low autocorrelation are desired, such as radar detection~\cite{beenker1985binary}, enhancement of stream-cipher key security~\cite{cai2009binary}, and improvement of impulse-response measurement accuracy in oil-well exploration~\cite{song2018analysis}.
It has also been extensively studied as a benchmark for quantum optimization~\cite{sanders2020compilation,bauer2024combinatorial,sciorilli2025competitive}, including the Quantum Approximate Optimization Algorithm (QAOA)~\cite{shaydulin2024evidence}, and the factorization machine with annealing (FMA)~\cite{nakano2025optimization}.
When the term $C_k(\bm{s})^2$ is expanded, interaction terms up to the fourth order, such as $s_i s_{i+k} s_j s_{j+k}$ $(i \neq j)$, appear in the energy $E(\bm{s})$.
Therefore, the LABS problem can naturally be formulated as a PUBO problem containing up to fourth-order terms.

For evaluation, the merit factor,
\begin{equation}
    F(\bm{s})=\frac{N^2}{2E(\bm{s})},
\end{equation}
is often used because it compensates for scaling with respect to the sequence length $N$ and enables fair comparison across different problem sizes~\cite{golay1982merit,golay1990new}.
However, since the primary objective of this study is to compare the performance of two formulations for the same sequence length $N$, cross-size comparison is not required.
Instead, we employ a normalized energy metric relative to the best-known minimum energy value $E_N^*$ for each sequence length, defined as
\begin{equation}
    \tilde{E}(\bm{s}) = \frac{E(\bm{s})}{E_N^*}.
\end{equation}
A value of $\tilde{E}(\bm{s}) = 1$ indicates that the best-known solution has been obtained, and this normalization provides an effective quantitative measure for comparing the relative performance of the PUBO and QUBO formulations.

\subsection{Vehicle routing problem with distance-balancing}
\label{sec:vrp_prob}

In this section, we first describe the basic structure of the vehicle routing problem (VRP) and then extend it to the distance-balancing variant investigated in this study.

The VRP was originally proposed by Dantzig and Ramser~\cite{dantzig1959truck} as a fundamental problem in network design and vehicle routing optimization.
\begin{figure}
    \centering
    \includegraphics[scale=0.4]{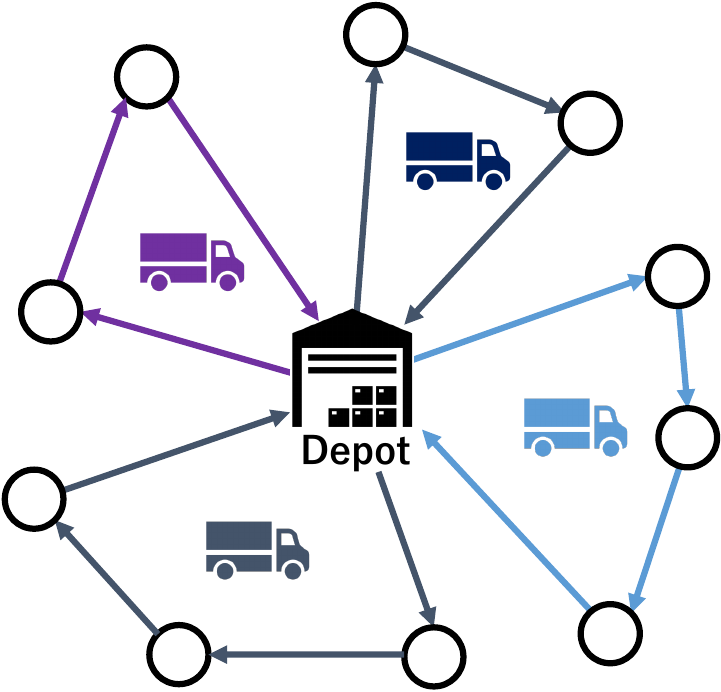}
    \caption{Conceptual illustration of the vehicle routing problem. In this example, one central depot is connected to ten distinct customer locations.
    The task is to design delivery routes such that all customer locations are visited, starting and ending at the depot. 
    Here, an example using four vehicles to complete the deliveries is shown.}
    \label{Fig:vrpimage}
\end{figure}
As illustrated in Fig.~\ref{Fig:vrpimage}, a single depot serves as both the origin and destination, and multiple vehicles are assigned to visit distinct customer locations.
Each vehicle departs from the depot, visits a subset of customer locations, and then returns to the depot.

In this study, we focus on an extended version of the VRP that aims to balance the travel distances among vehicles in addition to minimizing the total travel distance.
We choose this problem because it has both theoretical and practical significance.
Theoretically, the formulation represents a multi-objective optimization problem of the mean-variance type that naturally includes higher-order terms. 
It is particularly interesting because the structural characteristics of the higher-order interactions become mathematically explicit.
The objective function consists of two components, the first minimizes the total travel distance, and the second minimizes the variance of the travel distances among vehicles.
Since the polynomial orders of the two objectives are clearly separated, analyzing their respective results provides new insight into the effects of higher-order terms in multi-objective optimization.
Practically, minimizing the variance of travel distances helps reduce workload imbalances among vehicles, rather than solely minimizing total distance.
This approach contributes to fairer working hours and can shorten the overall completion time of deliveries.
In modern logistics operations, improving efficiency while maintaining better working conditions has become increasingly important, and the concept of distance-balancing directly reflects this social demand.

Let $P$ denote the number of customer locations (excluding the depot), $V$ the number of vehicles, and $S$ the number of travel steps in total.
As mentioned earlier, the problem considered in this study has two objectives, one representing the total travel distance and the other representing the variance of the travel distances among vehicles.
The problem also involves three constraints.
First, at each step, each vehicle can visit only one location.
Second, each location must be visited exactly once.
Third, once a vehicle returns to the depot, it stays there for the rest of the steps.

These objectives and constraints can be formulated using binary variables.
We define $x_{i,s}^{(v)} \in \{0,1\}$, which takes the value $1$ if vehicle $v$ visits location $i$ at step $s$, and $0$ otherwise.
Here, $i=0$ represents the depot, and $1 \le i \le P$ represent the customer locations.
Given the distance $d_{i,j}$ between locations $i$ and $j$, the two objective functions $f_1(\bm x)$ and $f_2(\bm x)$ are expressed as
\begin{align}
    \label{Eq:objective_1}
    f_1(\bm{x})
    &= \sum_{v=1}^V \sum_{i=1}^P d_{0,i}\, x_{i,1}^{(v)} \notag\\
    &\quad + \sum_{v=1}^V \sum_{s=1}^{S} \sum_{i=0}^P \sum_{j=0}^P
    d_{i,j}\, x_{i,s}^{(v)} x_{j,s+1}^{(v)}, \\[2pt]
    \label{Eq:objective_2}
    f_2(\bm{x})
    &= \frac{1}{V} \sum_{v=1}^V \Bigg[
      \sum_{i=1}^P d_{0,i}\, x_{i,1}^{(v)} \notag\\
      &+ \sum_{s=1}^{S} \sum_{i=0}^P \sum_{j=0}^P
        d_{i,j}\, x_{i,s}^{(v)} x_{j,s+1}^{(v)}
      - \frac{f_1(\bm{x})}{V}
    \Bigg]^2, \\
    \bm x &= \{x_{i, s}^{(v)} | 0 \le i \le N, 1 \le s \le S, 1 \le v \le V\}.
\end{align}
The number of steps $S$ is an integer satisfying $S \le P$.

The three constraints can be represented using penalty functions $g_1(\bm{x})$, $g_2(\bm{x})$, and $g_3(\bm{x})$, each taking a value of zero when the corresponding constraint is satisfied:
\begin{align}
    \label{Eq:constrained_1}
    &g_1({\bm x}) = \sum_{s=1}^S \sum_{v=1}^V \left[ 
    \sum_{i=1}^P x_{i,s}^{(v)} - 1
    \right]^2,\\
    \label{Eq:constrained_2}
    &g_2({\bm x}) = \sum_{i=1}^P \left[
    \sum_{v=1}^V \sum_{s=1}^S x_{i,s}^{(v)} - 1
    \right]^2,\\
    \label{Eq:constrained_3}
    &g_3({\bm x}) = \sum_{v=1}^V \sum_{s=1}^{S-1} x_{0,s}^{(v)} (1-x_{0,s+1}^{(v)}).
\end{align}

Based on these definitions, the overall energy function for the vehicle routing problem with distance-balancing can be written as 
\begin{align}
    \label{Eq:energyfunction_original}
    E({\bm x}) = &(1-\alpha)f_1({\bm x}) + \alpha f_2({\bm x}) \notag\\
                 &+ \mu (g_1({\bm x}) + g_2({\bm x}) + g_3({\bm x})),
\end{align}
where $\alpha$ is a hyperparameter controlling the weight of the variance term, and $\mu$ is a hyperparameter representing the strength of the constraints.
It should be noted that, due to the squared term in \eqref{Eq:objective_2}, the energy function in \eqref{Eq:energyfunction_original} contains third- and fourth-order terms, and therefore cannot be represented in the QUBO form defined in \eqref{Eq:QUBO}.
\section{Numerical results}
\label{sec:results}

\subsection{Experimental setup}
\label{sec:setup}
All numerical experiments in this study were conducted using Amplify AE.
As described in Section~\ref{sec:background}, Amplify AE implements both a conventional QUBO solver and a PUBO solver capable of directly handling up to fourth-order terms, thereby enabling a fair comparison between the two formulations on the same platform.
In the experiments, the PUBO solver directly optimized the original higher-order objective functions, while the QUBO solver employed the substitution method to perform order reduction prior to computation. 
The penalty coefficient in the substitution method was fixed to the default value of $p=1$ provided by the Amplify system for all experiments.
Both solvers were executed on identical problem instances under the same computational conditions.
The computation time was set to 60 seconds for both problems.
All other internal parameters of Amplify AE were kept at their default values throughout the experiments.

\subsection{LABS results}
\label{sec:labs_results}
In this section, we present the results of comparing the solution accuracy between the PUBO and QUBO solvers for the LABS problem.
The sequence length $N$ was varied in the range $5 \le N \le 99$, and ten independent trials were performed for each $N$.
The computation time was limited to 60 seconds, and both solvers were executed under identical experimental conditions.

The evaluation metric used in this analysis is the normalized energy introduced in Section~\ref{sec:labs_prob}, which is defined as
\begin{equation}
    \tilde{E}(\bm{s}) = \frac{E(\bm{s})}{E_N^*},
\end{equation}
where $E_N^*$ denotes the best-known optimal energy for a sequence length $N$.
A value of $\tilde{E}=1$ indicates that the optimal solution was obtained.

\begin{figure}[t]
    \centering
    \includegraphics[scale=0.3]{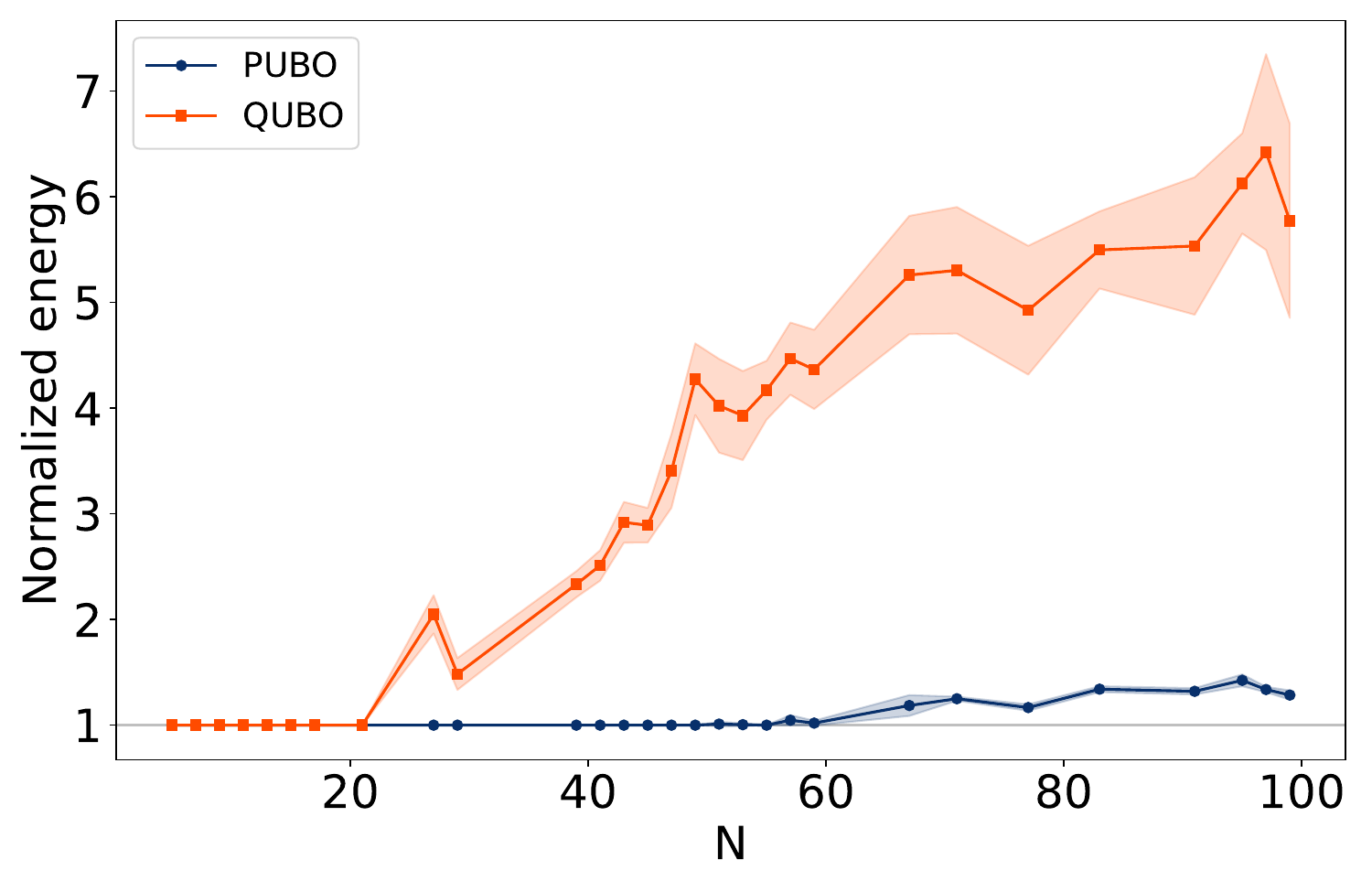}
    \caption{Relationship between the sequence length $N$ and average normalized energy $\tilde{E}$ in the LABS problems. 
    The plot shows the mean of ten runs, and the shaded area represents the standard deviation.}
    \label{Fig:LABSresult}
\end{figure}
Fig.~\ref{Fig:LABSresult} shows the average normalized energy $\tilde{E}$ obtained from ten independent trials for each sequence length $N$.
The shaded area represents the standard deviation over the trials.
The results show that the QUBO solver exhibited degraded solution accuracy even for relatively small $N$, accompanied by large variations in the obtained results.
In contrast, the PUBO solver consistently achieved lower average $\tilde{E}$ values with smaller deviations, indicating more stable and higher-quality solutions.
Furthermore, the performance gap between PUBO and QUBO increased with problem size, clearly demonstrating the scalability advantage of directly handling higher-order terms in the PUBO formulation.

\subsection{VRP results}
\label{sec:vrp_results}
\begin{figure*}
  \centering
  \begin{minipage}[b]{0.32\textwidth}
    \centering
    \includegraphics[keepaspectratio,width=\linewidth]{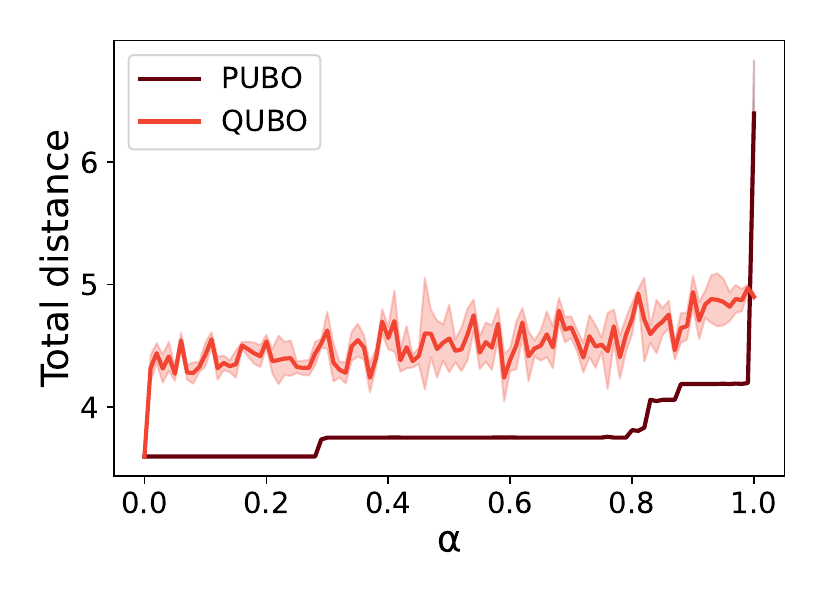}
    (a)
  \end{minipage}\hfill
  \begin{minipage}[b]{0.32\textwidth}
    \centering
    \includegraphics[keepaspectratio,width=\linewidth]{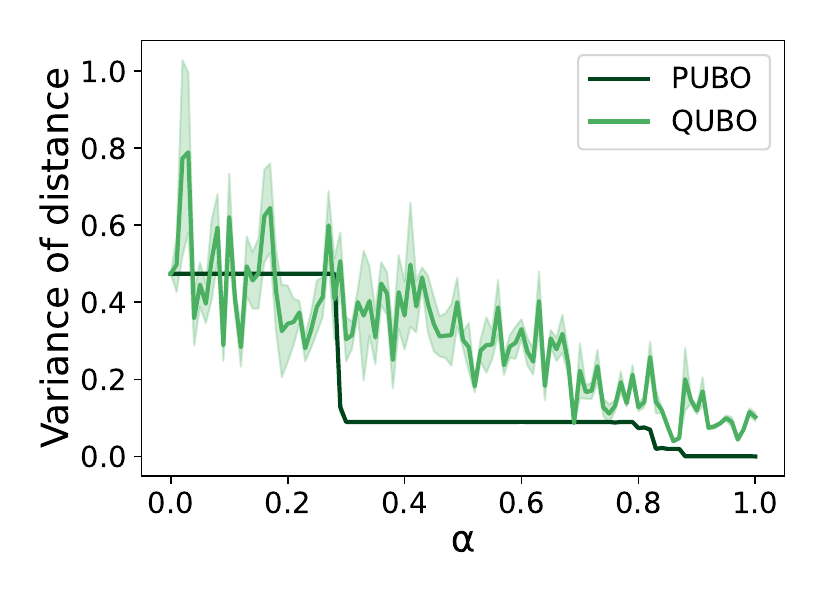}
    (b)
  \end{minipage}\hfill
  \begin{minipage}[b]{0.32\textwidth}
    \centering
    \includegraphics[keepaspectratio,width=\linewidth]{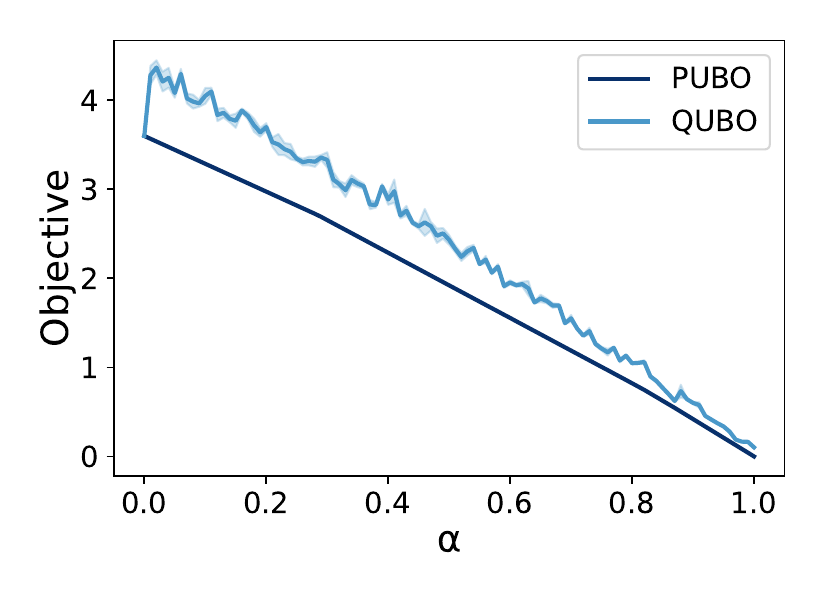}
    (c)
  \end{minipage}
  \caption{Results for a single VRP instance with $P=11$ customer locations and $V=3$ vehicles.
  Shown are the relationships between the hyperparameter $\alpha$ and (a) total travel distance, (b) variance of travel distances, (c) objective function value. 
  Each curve represents the mean of ten independent runs with a runtime of 60 seconds, and the shaded areas represent standard deviations.}
  \label{Fig:VRPresult_single}
\end{figure*}
In this section, we present the numerical results for the vehicle routing problem with distance-balancing introduced in Section~\ref{sec:vrp_prob}.
The analysis was conducted in two stages.
First, for a single problem instance, the hyperparameter $\alpha$ controlling the weight of the variance term was varied, and the total travel distance, the variance of travel distances, and the overall objective function values were compared.
In addition, scatter plots of all obtained solutions were examined to analyze the distribution characteristics of the solutions.
Second, for multiple randomly generated problem instances, the performance of the direct PUBO formulation and the QUBO formulation obtained through order reduction was quantitatively compared using the hypervolume indicator, which measures the diversity and quality of the non-dominated solution sets.

\subsubsection{Single-instance analysis}
\label{sec:vrp_results_single}
In the numerical experiments, the numbers of customer locations and vehicles were set to $P=11$ and $V=3$, respectively, with the locations being uniformly distributed within a unit square.
In practical delivery scenarios, multiple vehicles share the visiting tasks, therefore, setting the total number of steps $S$ equal to the number of locations is redundant.
To maintain search flexibility while avoiding unnecessary duplication, the number of steps $S$ was defined as follows:
\begin{equation}
    S=\left\lceil \frac{P}{\,V-1\,} \right\rceil.
    \label{Eq:VRP_step}
\end{equation}
The penalty coefficients representing the strengths of the three constraint terms were set to identical values.
Let $\max_{i,j} d_{i,j}$ denote the maximum distance between any two locations.
The penalty coefficient was then defined as
\begin{equation}
    \mu=\max_{i,j}d_{i,j}+(\max_{i,j}d_{i,j}/{V})^2.
    \label{Eq:VRP_mu}
\end{equation}

Based on \eqref{Eq:energyfunction_original}, the hyperparameter $\alpha$ was varied from 0 to 1 in increments of $0.01$ to investigate the trade-off between the two objective functions, namely the total travel distance and the variance of travel distances.
Ten independent trials were performed for each value of $\alpha$, and the computation time was fixed at 60 seconds.

Fig.~\ref{Fig:VRPresult_single} shows the relationships among $\alpha$, the total travel distance, the variance of travel distances, and the objective function values.
The solid lines represent the averages of ten independent trials, and the surrounding shaded regions indicate the standard deviations.
As shown in Fig.~\ref{Fig:VRPresult_single}(a), the total travel distance tended to increase with larger values of $\alpha$.
This behavior occurs because prioritizing the reduction of distance variance leads to route selections that include slightly longer detours.
In the PUBO results, this trend appeared consistently with small variations among trials, whereas in the QUBO results, larger fluctuations were observed, indicating less stability in solution quality.
As shown in Fig.~\ref{Fig:VRPresult_single}(b), the variance of travel distances decreased with increasing $\alpha$ in the PUBO results, indicating better performance than the QUBO solver.
In addition, as shown in Fig.~\ref{Fig:VRPresult_single}(c), the objective function values obtained by the PUBO solver were consistently lower than those obtained by the QUBO solver, demonstrating that the PUBO solver provides stable and high-quality solutions.

\begin{figure}
    \centering
    \includegraphics[scale=0.3]{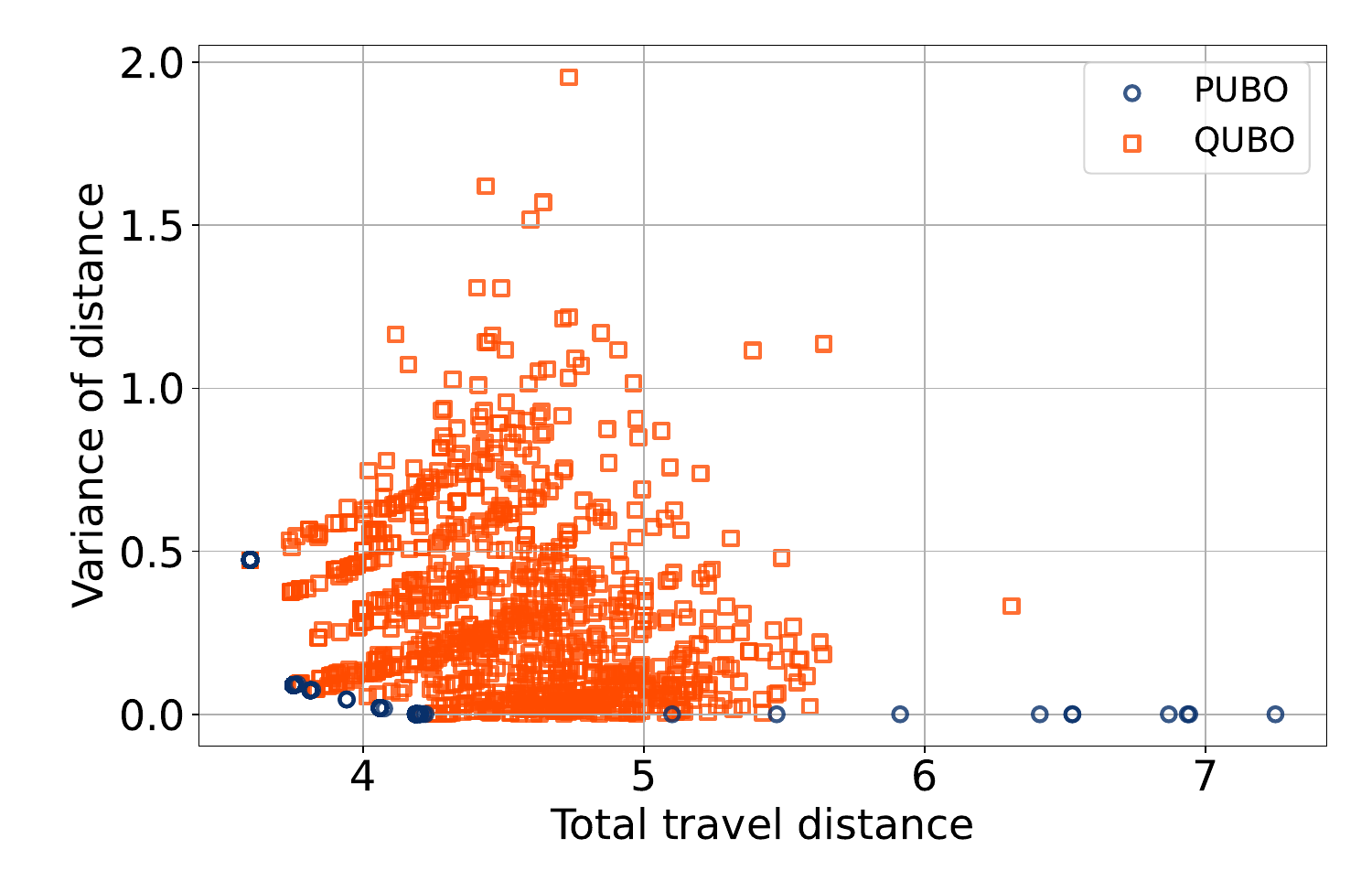}
    \caption{Scatter plot showing the trade-off between the two objective values. The x-axis represents the total travel distance and the y-axis represents the variance of travel distances among vehicles. The results are obtained from 10 independent runs for each of 100 different values of $\alpha$. Only feasible solutions that satisfy all constraints are shown, while infeasible ones are omitted.}
    \label{Fig:VRP_distanceVSvariance}
\end{figure}
Fig.~\ref{Fig:VRP_distanceVSvariance} shows a scatter plot of all feasible solutions obtained over all values of $\alpha$ and all trials, where the horizontal axis represents the total travel distance and the vertical axis represents the variance of travel distances.
In the scatter plot, the solutions obtained by the PUBO solver were concentrated near the Pareto front, accurately reflecting the trade-off between the two objectives.
In contrast, the solutions obtained by the QUBO solver were widely scattered, and several solutions exhibited large values in both objective functions.
While the QUBO solver generated some solutions with moderately small variance, it was unable to reach the true minimum region along the vertical axis. 
This limitation reflects the structural difference between the two objectives: the total distance, expressed by quadratic terms, is well captured, whereas the variance component includes fourth-order interactions that the QUBO formulation cannot faithfully express. 
As a result, the variance-minimizing capability of QUBO remained intrinsically constrained compared to the direct PUBO formulation.

\subsubsection{Multi-instance hypervolume analysis}
\label{sec:vrp_results_multi}
In Section~\ref{sec:vrp_results_single}, we demonstrated through single-instance analysis that directly retaining higher-order terms in the PUBO formulation enables more stable and accurate solutions than those obtained after order reduction.
In this section, we quantitatively evaluate the advantage of PUBO, which is unaffected by order-reduction transformations, from the perspective of multi-objective optimization using multiple problem instances.

The number of customer locations excluding the depot was fixed at $P=11$, and the number of vehicles was fixed at $V=3$.
30 problem instances were generated by uniformly placing the locations within a unit square.
All other parameters were identical to those used in Section~\ref{sec:vrp_results_single}.

For a quantitative evaluation of the solution sets, we employed the hypervolume indicator, which is widely used in multi-objective optimization~\cite{eckart1999multiobjective,andreia2021hypervolume}.
The hypervolume is defined as the volume of the region dominated by the non-dominated solution set relative to a reference point.
A larger hypervolume value indicates greater diversity and coverage within the obtained Pareto-optimal solutions.
In this study, to ensure consistency across all instances, the reference point was set to the maximum observed value of each objective function plus a margin of 0.1.

\begin{figure}
    \centering
    \includegraphics[scale=0.35]{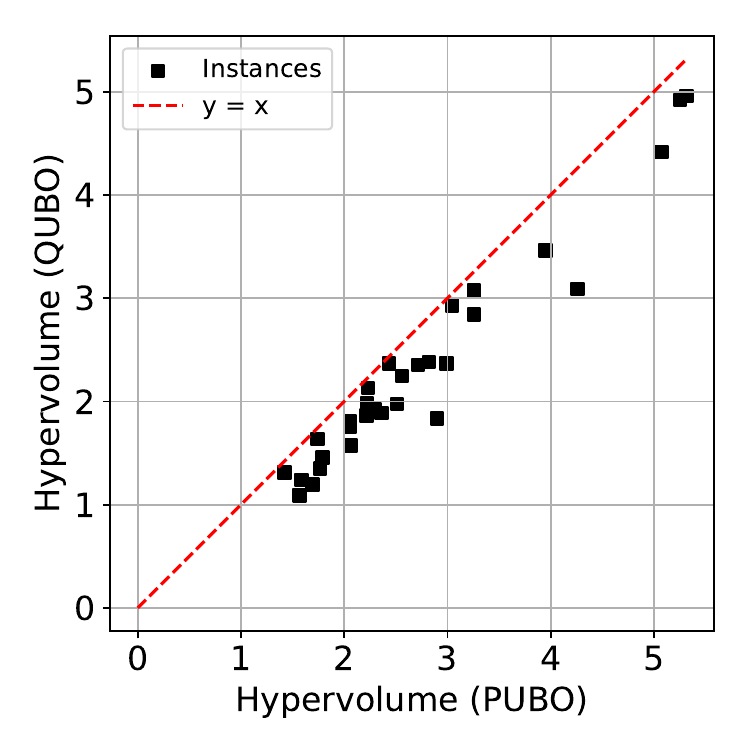}
    \caption{Comparison of hypervolume values between the PUBO and QUBO solvers across 30 randomly generated VRP instances. 
    The x-axis shows the hypervolume values obtained by the PUBO solver and the y-axis shows that obtained by the QUBO solver. 
    Each point corresponds to one instance, and the dashed red line indicates $y=x$.}
    \label{Fig:VRPresult_hypervolume}
\end{figure}
Fig.~\ref{Fig:VRPresult_hypervolume} shows the comparison of hypervolume values obtained by the PUBO and QUBO solvers across 30 problem instances.
The horizontal axis represents the hypervolume values obtained by the PUBO solver, and the vertical axis represents those obtained by the QUBO solver.
The red dashed line indicates $y=x$.
Each point corresponds to one problem instance, and points below the dashed line indicate instances where the PUBO solver achieved larger hypervolume values than the QUBO solver.

As a result, all instances were distributed below the $y=x$ line, confirming that the PUBO solver consistently achieved larger hypervolume values than the QUBO solver.
This demonstrates that the PUBO solver can faithfully reproduce the trade-off structure between the total travel distance and the variance of travel distances while satisfying the constraint conditions defined in \eqref{Eq:energyfunction_original}, thereby generating stable and high-quality solution sets in multi-objective optimization.
In contrast, in the QUBO solver, the introduction of auxiliary variables and penalty terms likely increased the complexity of the search space, thereby degrading the trade-off relationships among the obtained solutions.

Overall, these results, ranging from detailed single-instance analysis to quantitative evaluation across multiple instances, clearly demonstrate the intrinsic advantage of the PUBO formulation, which directly handles higher-order terms without order reduction.
\section{Discussion}
\label{sec:discussion}

\subsection{Comparison of the number of variables between PUBO and QUBO}
One of the most evident drawbacks of order reduction is the increase in the number of variables resulting from the introduction of auxiliary variables.
In both the LABS problem and the vehicle routing problem with distance-balancing considered in this study, the QUBO formulation requires a large number of auxiliary variables to account for higher-order terms.
\begin{figure}[t]
  \begin{minipage}[b]{0.48\linewidth}
    \centering
    \includegraphics[keepaspectratio, scale=0.265]{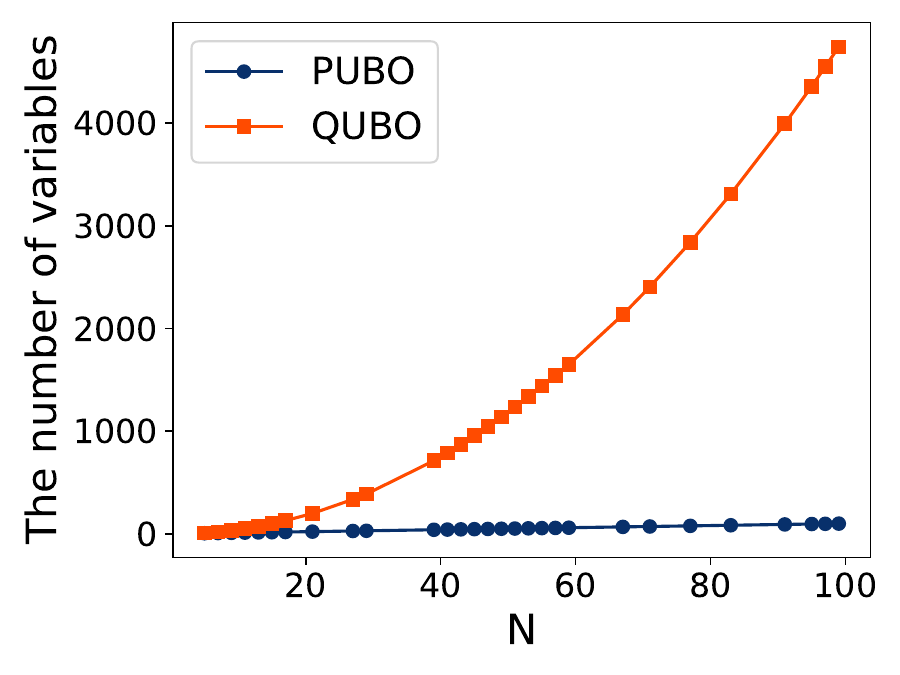}
    (a)
  \end{minipage}
  \begin{minipage}[b]{0.48\linewidth}
    \centering
    \includegraphics[keepaspectratio, scale=0.265]{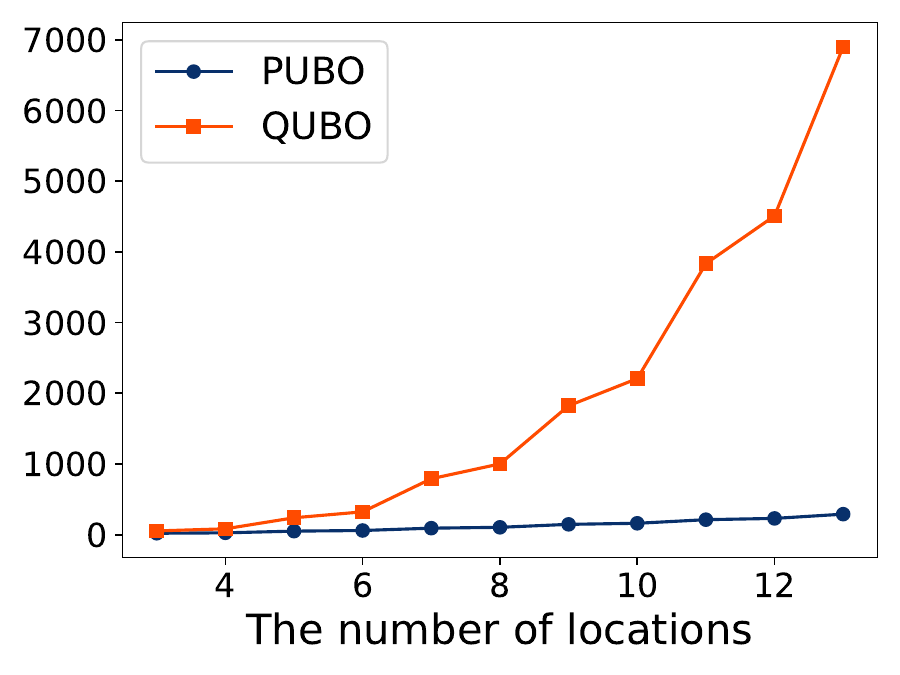}
    (b)
  \end{minipage}
  \caption{Comparison of the number of binary variables between the PUBO formulation (blue) and the QUBO formulation after order reduction (orange). 
  (a) LABS problem as a function of the sequence length $N$.
  (b) VRP as a function of the number of customer locations $P$.}
  \label{Fig:var_comparison}
\end{figure}

Fig.~\ref{Fig:var_comparison} illustrates the increase in the total number of variables as a function of problem size.
Fig.~\ref{Fig:var_comparison}(a) shows the relationship between the sequence length $N$ and the number of variables in the LABS problem.
As observed in Fig.~\ref{Fig:var_comparison}(a), the difference between the PUBO and QUBO formulations becomes increasingly pronounced as the problem size increases.
This is because the number of auxiliary variables introduced during order reduction increases rapidly with the number of higher-order terms in the problem.
In particular, for large-scale problems, the number of variables after order reduction reaches several tens of times that in the direct PUBO formulation, clearly indicating that the effect of order reduction becomes dominant.

A similar trend was observed for the VRP, as shown in Fig.~\ref{Fig:var_comparison}(b).
In this case, the number of vehicles was fixed at three, and the number of customer locations $P$ was varied.
As the number of locations increased, the total number of variables also increased in both formulations, however, the influence of auxiliary variables introduced by order reduction became increasingly significant for larger $P$, leading to a rapid widening of the gap between PUBO and QUBO.

These results indicate that the impact of order reduction becomes increasingly significant for larger problem sizes and is a major factor contributing to model expansion.
This increase in the number of variables expands the search space, resulting in higher computational costs and reduced solution stability.
In contrast, directly handling higher-order terms in the PUBO formulation avoids this increase in variables and enables stable generation of high-quality solutions.

\subsection{Impact of penalty coefficient of order reduction on solution quality}
When performing order reduction, as described in Section~\ref{sec:order_reduction}, a penalty term is introduced to satisfy the substitution constraint, and the corresponding penalty coefficient must be properly determined.
This coefficient is a crucial parameter that balances the strictness of the constraint with the flexibility of the search process, and its value significantly affects the solution quality.
Although there is likely an optimal penalty coefficient for each problem size, how this optimal value changes with problem scale remains unclear.
Therefore, in this section, we investigate how the optimal penalty coefficient depends on the sequence length $N$ using the LABS problem as a test case.
The computation time was fixed at 60 seconds, consistent with the settings described in Section~\ref{sec:results}, and the penalty coefficient $p$ for order reduction was varied within the range $1.0 \leq p \leq 10$.
The problem size was set in the range $5 \leq N \leq 99$.

\begin{figure}
    \centering
    \includegraphics[scale=0.32]{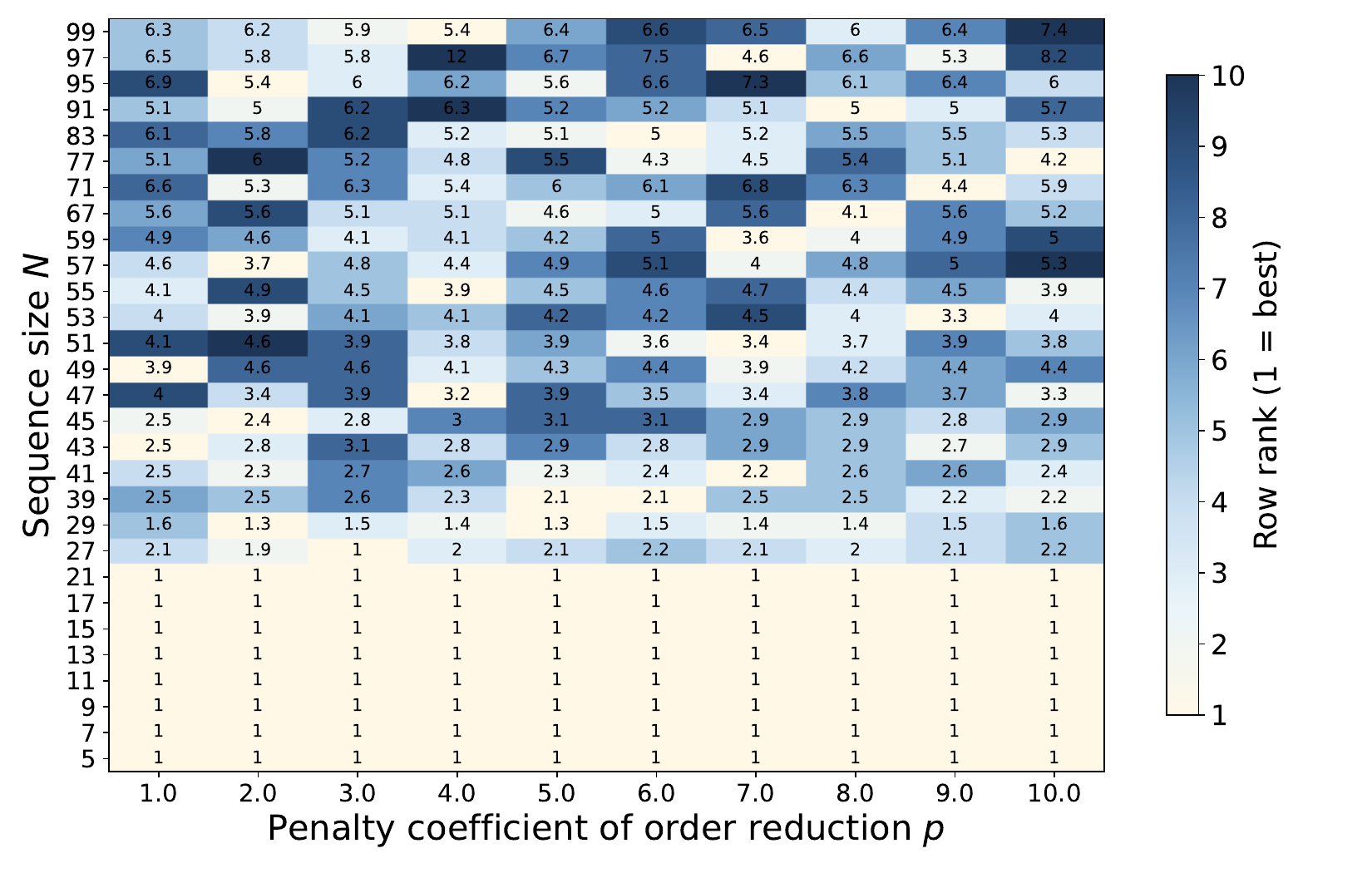}
    \caption{Heatmap visualization of solution quality with respect to the penalty coefficient and problem size in the LABS problem.
    The horizontal axis represents the penalty coefficient $p$ used in the order-reduction method, and the vertical axis represents the sequence length $N$.
    The color indicates the relative ranking of the obtained objective values for each problem size, with lighter colors corresponding to relatively better results.}
    \label{Fig:LABS_heatmap}
\end{figure}
Fig.~\ref{Fig:LABS_heatmap} shows a heatmap where the horizontal axis represents the penalty coefficient used for order reduction, the vertical axis represents the sequence length $N$, and the color indicates the ranking of solutions obtained under each condition.
Brighter colors correspond to smaller energy values, indicating better solutions.
As observed in Fig.~\ref{Fig:LABS_heatmap}, the optimal penalty coefficient varies depending on the problem size, and no single value provides consistently good performance across all $N$.
Therefore, determining an appropriate penalty coefficient is not straightforward and requires problem-specific tuning.
\section{Conclusion}
\label{sec:conclusion}
This study investigated the effectiveness of directly solving combinatorial optimization problems that include higher-order terms without applying order reduction.
As benchmarks, we employed the low autocorrelation binary sequence problem, which naturally involves fourth-order interactions, and the vehicle routing problem with distance-balancing.
Comparisons between the PUBO and QUBO formulations were performed using the Fixstars Amplify Annealing Engine.
The numerical results demonstrated that the PUBO formulation, which directly handles higher-order terms, suppresses the exponential increase in the number of variables that arise in QUBO formulations due to order reduction, while maintaining stable and efficient solution search performance.
In particular, for large-scale problems, the QUBO formulation introduced a large number of auxiliary variables, expanding the search space and resulting in unstable and less accurate solutions.
In contrast, the PUBO formulation preserved the original problem structure and achieved more consistent and higher-quality solutions.
Furthermore, the comparison of multi-objective optimization results for the VRP confirmed that the direct PUBO formulation successfully reproduced a broader trade-off structure and achieved greater diversity in the obtained solution sets.
Overall, these findings indicate that the direct PUBO formulation offers a promising approach that achieves both structural simplicity and computational stability compared with QUBO formulations requiring order reduction.
Future work will focus on extending the applicability of this approach to larger and more complex real-world problems and on collaborating with the development of hardware and solvers capable of directly executing PUBO models, thereby further promoting the practical application of higher-order combinatorial optimization.

\section*{Acknowledgment}
S. Tanaka wishes to express their gratitude to the World Premier International Research Center Initiative (WPI), MEXT, Japan, for their support of the Human Biology-Microbiome-Quantum Research Center (Bio2Q).

\bibliography{reference.bib}

\begin{IEEEbiography}[{\includegraphics[width=1in,height=1.25in,clip,keepaspectratio]{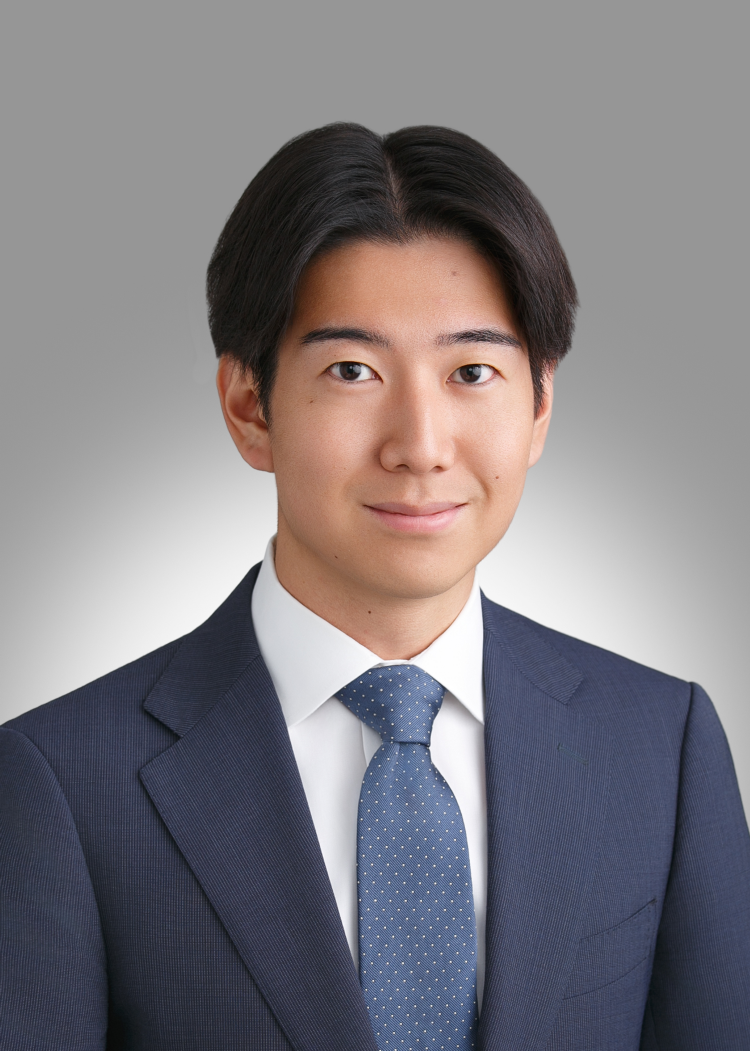}}]{Kazuki Ikeuchi} 
    received the B. Eng. degree in applied physics and physico-informatics from Keio University, Kanagawa, Japan, in 2023, where he is currently pursuing the M. Eng. degree in fundamental science and technology. 
    His research interests include mathematical optimization, quantum annealing, Ising machines, and practical applications of quantum-inspired technologies in industry.
\end{IEEEbiography}

\begin{IEEEbiography}[{\includegraphics[width=1in,height=1.25in,clip,keepaspectratio]{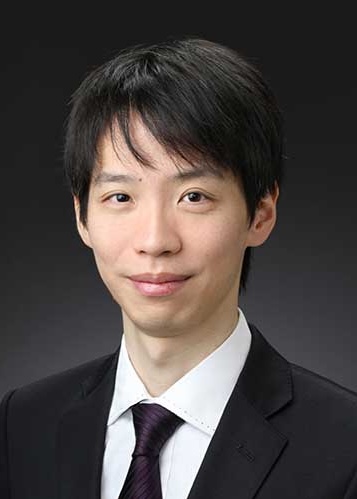}}]{Yoshiki Matsuda} 
    received his B. Sci. degree in 2007, M. Sci. degree in 2009, and Dr. Sci. in 2011 from the Tokyo Institute of Technology (now the Institute of Science Tokyo), Japan. He is currently the COO of Fixstars Corporation and the CEO of Fixstars Amplify Corporation. His research interests include HPC, GPU calculation, quantum annealing, Ising machines, quantum computing, and statistical mechanics.
\end{IEEEbiography}

\begin{IEEEbiography}[{\includegraphics[width=1in,height=1.25in,clip,keepaspectratio]{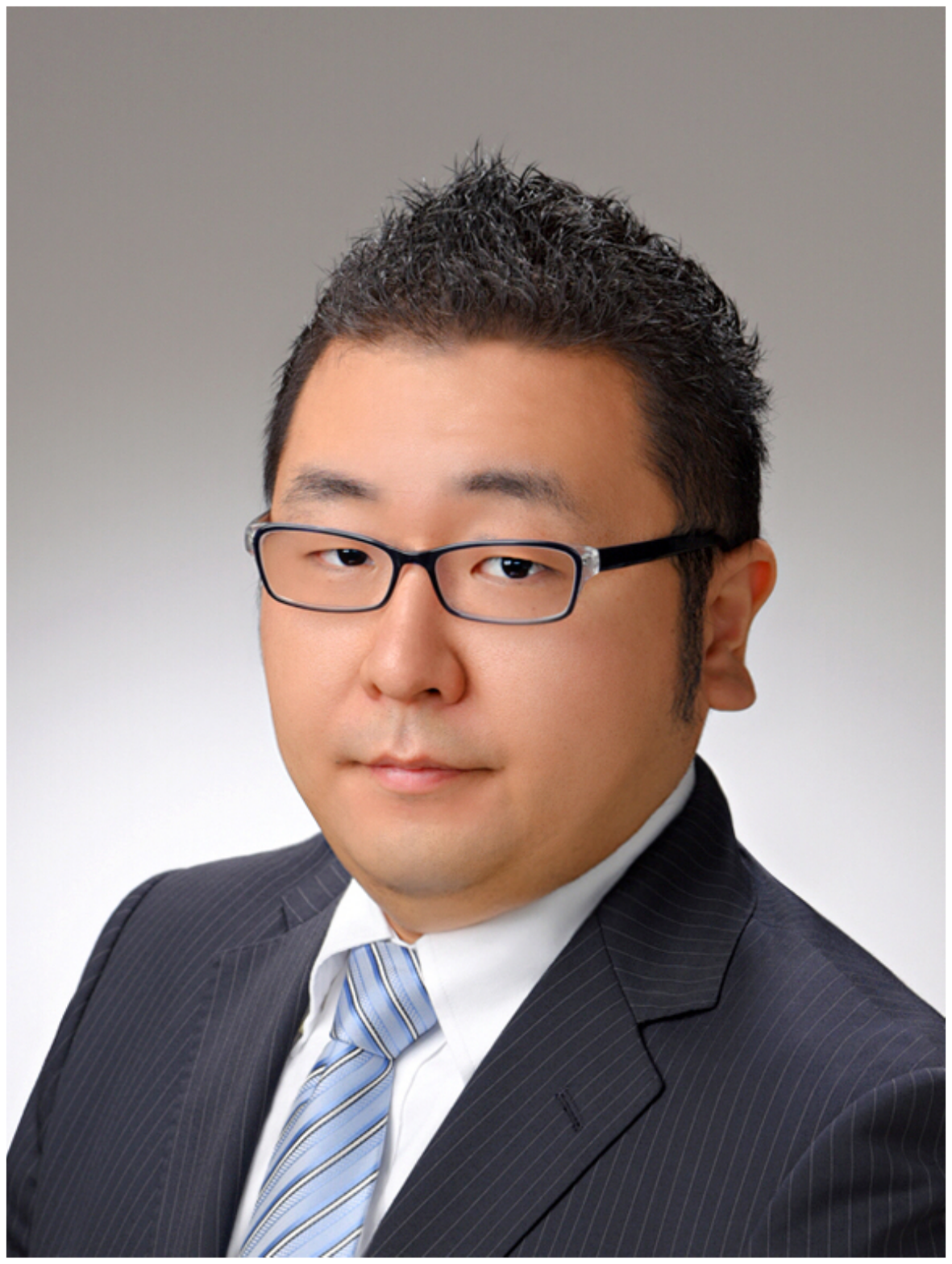}}]{Shu Tanaka} 
    (Member, IEEE) received the B. Sci. degree from the Tokyo Institute of Technology, Tokyo, Japan, in 2003, and the M. Sci.
    and Dr. Sci. degrees from the University of Tokyo, Tokyo, Japan, in 2005 and 2008, respectively. 
    He is currently a Professor in the Department of Applied Physics and Physico-Informatics, Keio University, a chair of the Keio University Sustainable Quantum Artificial Intelligence Center (KSQAIC), Keio University, a Core Director at the Human Biology-Microbiome-Quantum Research Center (Bio2Q), Keio University, and a Guest Senior Researcher (Guest Professor), Green Computing Systems Research Organization, Waseda University.
    His research interests include quantum annealing, Ising machines, quantum
    computing, statistical mechanics, and materials science. 
    He is a member of the Physical Society of Japan (JPS), and the Information Processing Society
    of Japan (IPSJ).
\end{IEEEbiography}

\EOD

\end{document}